\documentclass{elsarticle}
\usepackage{graphicx}%
\usepackage{multirow}%
\usepackage{amsmath,amssymb,amsfonts}%
\usepackage{amsthm}%
\usepackage{mathrsfs}%
\usepackage[title]{appendix}%
\usepackage{xcolor}%
\usepackage{textcomp}%
\usepackage{manyfoot}%
\usepackage{booktabs}%
\usepackage{algorithm}%
\usepackage{algorithmicx}%
\usepackage{algpseudocode}%
\usepackage{listings}%
\usepackage{verbatim}
\usepackage{enumitem}
 \usepackage{comment}
 \usepackage{amsmath}

\journal{}
\begin{document}

\begin{frontmatter}

\title{Revisiting the non-equilibrium phase transitions of the continuous-trait Axelrod model}

\author[GMU]{Sandro M. Reia}
%\ead{smarti71@gmu.edu }

\author[UFPE]{Paulo R.  A. Campos}
%\ead{paulo.acampos@ufpe.br}

\author[mymainaddress]{Jos\'e F.  Fontanari}
%\ead{fontanari@ifsc.usp.br}

\address[GMU]{Geography and Geoinformation Science, College of Science, George Mason University, Fairfax, VA, USA}

\address[UFPE]{Departamento de F\'{\i}sica, Centro de Ci\^encias Exatas e da Natureza,  Universidade Federal de Pernambuco,  50740-560 Recife,  Pernambuco,  Brazil}

\address[mymainaddress]{Instituto de F\'{\i}sica de S\~ao Carlos,  Universidade de S\~ao Paulo,  13566-590 S\~ao Carlos,  S\~ao Paulo,  Brazil}

\begin{abstract}
We investigate the non-equilibrium phase transitions of the continuous-trait Axelrod model, an agent-based framework where individual culture is represented by a vector of $F$ continuous features confined to the interval $(0,1)$. Local interactions are governed by a metric similarity threshold $d$, which acts as a continuous control parameter of social tolerance. The dynamics inevitably freeze into one of two absorbing configuration classes: an ordered, homogeneous monocultural state at high tolerance, or a highly fragmented, disordered state at low tolerance. While previous studies characterized the transition as hybrid based on the continuous behavior of the domain density $\mu$ alongside a discontinuous jump in the largest domain fraction $\rho$,  we show that this apparent continuity is an artifact of severe finite-size masking effects. By shifting the methodological focus to the scaling of the median $\tilde{\mu}$ and analyzing the full probability distributions $P(\mu)$, we unveil a clear bimodal structure with disjoint maxima across independent simulation runs. Our results reveal that for $F=2$, the system undergoes a genuinely hybrid transition in the contemporary sense, featuring a tiny but finite latent jump ($\mu_c \approx 0.089$) at the critical  threshold  $d_c \approx 0.0784$ while scaling toward it from below via a non-analytical power law with a mean-field exponent $\beta \approx 1/2$. Conversely, for $F=3$, the higher trait-space dimensionality suppresses local fluctuations, yielding a traditional, non-hybrid first-order transition. We apply this  framework to the alternative discrete Poisson variant of the model, successfully confirming its known continuous transition for $F=2$ and discontinuous, non-hybrid transition for $F=3$, thereby establishing a unified characterization of phase transitions in Axelrod-like systems.
\end{abstract}

%\keywords{synchronization, collective motion, correlated random-walk, spatial accessibility}



\end{frontmatter}

\newpage

\section{Introduction}\label{sec:intro}

Social influence and homophily---the tendency of individuals to interact preferentially with similar others---are key drivers in the preservation of cultural diversity~\cite{McPherson_2001, Marsden_1993}. The study of social influence, consensus, and fragmentation has grown into a mature branch of statistical mechanics, providing valuable insights into complex societal dynamics (for a comprehensive review on modern social physics, see Jusup et al.~\cite{Jusup_2022}). A quantitative understanding of these factors advanced significantly with the agent-based model proposed by Robert Axelrod~\cite{Axelrod_1997}. In this model, agents are represented by a vector of $F$ cultural features, where each feature can adopt one of $Q$ distinct nominal traits. The interaction probability between two neighbors is proportional to their cultural overlap, while the interaction itself allows them to share an additional trait. This mechanism creates a positive feedback loop where similarity promotes interaction, and interaction further increases similarity. Ultimately, because agents with entirely different cultural profiles are strictly forbidden from interacting, this local convergence rule prevents complete assimilation, sustaining stable, multicultural absorbing states, provided that the initial cultural diversity $Q$ is sufficiently large~\cite{Axelrod_1997}.

This feedback mechanism drives a non-equilibrium phase transition that separates an ordered, monocultural phase at low values of $Q$ from a disordered, highly fragmented phase as $Q$ increases~\cite{Castellano_2000, Vilone_2002, Klemm_2003, Klemm_2003b}. Accurately classifying the nature of such absorbing-state transitions is a central challenge. In the modern statistical-physics sense, they can even manifest as genuinely hybrid phase transitions~\cite{Dorogovtsev_2006, Baxter_2015, Lee_2016}---a phenomenon uniquely characterized by the coexistence of a discontinuous jump in the order parameter, typical of first-order transitions, and a non-analytical power-law scaling as the critical point is approached. The non-equilibrium phase transition of Axelrod's model departs from the traditional non-equilibrium paradigm characterized by a fluctuating active phase collapsing into a unique absorbing state~\cite{Marro_1999}. Instead, the dynamics inevitably evolve toward a frozen steady state where all remaining configurations are absorbing---meaning that any pair of adjacent agents either shares identical traits or differs across all features, halting further evolution. The phase transition is therefore defined by the distinct nature of these frozen configurations. This characterization shares conceptual parallels with standard geometric percolation~\cite{Stauffer_1992}, albeit shaped by a history-dependent dynamical path rather than purely static states.

A major technical obstacle in investigating this transition stems from the discrete nature of the model's control parameters. Because both $F$ and $Q$ are integer variables, it is  impossible to explore the immediate neighborhood of the critical point or to apply standard finite-size scaling techniques to determine critical exponents~\cite{Privman_1990}. To bypass this constraint, Castellano et al.~\cite{Castellano_2000} introduced the Poisson variant, where the integer trait for each cultural feature is drawn from a Poisson distribution governed by a continuous parameter $q \in (0, \infty)$. Transitioning to $q$ allowed for a systematic exploration of the critical region and a more precise statistical mechanics characterization of the phase diagram. In particular, for $F=2$, the system undergoes a continuous transition at $q_c \approx 3.10$ for both order parameters---the density of cultural domains $\mu$~\cite{Peres_2015} and the fraction of agents in the largest domain $\rho$~\cite{Reia_2016}. For $F=3$, however, the transition was characterized as discontinuous based on the distribution of domain sizes~\cite{Castellano_2000}, rather than on the direct behavior of these order parameters.

Despite solving the parameter continuity problem, the Poisson variant introduces significant computational overhead, as the stochastically shifting number of traits across initial realizations demands massive ensemble averaging. To overcome this bottleneck and achieve a higher degree of behavioral realism, a generalized model that uses continuous cultural traits has emerged as a promising alternative~\cite{Campos_2025}. In this formulation, the features of an agent are continuous variables confined to the interval $(0,1)$, and interaction is governed by a metric similarity threshold, $d$~\cite{Mora_2025}. Neighbors can only interact if the distance between their traits is within $d$, which acts as a smooth control parameter. From a sociological perspective, this model provides a more realistic representation of human culture, where opinions, belief systems, and economic choices naturally evolve along spectrums rather than fitting into rigid slots~\cite{Buechel_2014, Fogarty_2024}. Moreover, this geometric approach allows the threshold $d$ to be interpreted as a proxy for social tolerance, establishing a transparent link between local rules and global outcomes.

In a recent study exploring this continuous model under perfect copying rules~\cite{Campos_2025}, the transition was preliminarily characterized as hybrid, albeit in a non-standard sense: while $\rho$ vanished abruptly via a discontinuous jump, the domain density $\mu$ was interpreted as a continuous order parameter, leading to a ``hybrid" classification based on the contrasting behavior of two distinct order parameters rather than a single field. However, determining the true order of a transition through Monte Carlo simulations without prior analytical guidance is a notoriously challenging problem, particularly for weak first-order transitions~\cite{Landau_2000}. A classic illustration is the two-dimensional Potts model, which exhibits a clear first-order transition for ten states, but a remarkably weak one for five states~\cite{Baxter_1982}, resembling a continuous transition in finite systems. In the continuous-trait Axelrod model, the number of features $F$ plays an analogous role to the number of states in the Potts model, adding degrees of freedom that sharpen the discontinuous nature of the transition.

In this work, we show that the apparent continuity of the domain density previously reported for $F=2$ was an artifact of severe finite-size masking effects. We focus our analysis specifically on the behavior of $\mu$, rather than on $\rho$, since the discontinuous nature of the latter is already unambiguously established by the distinct crossing of $\rho$ versus $d$ curves for different linear lattice sizes $L$~\cite{Campos_2025}. To unveil the true thermodynamic behavior of $\mu$, we shift our methodological approach from standard ensemble averages to the scaling of the median $\tilde{\mu}$, and significantly expand the simulation scope to large lattice sizes. It is worth emphasizing that relying on the median rather than the ensemble mean is particularly crucial for absorbing-state models where the probability distributions become strongly skewed or clearly bimodal near the transition threshold. In such scenarios, the mean fails to represent the typical macroscopic behavior of the system, whereas the median robustly captures the true structural state.

The clearest confirmation of the discontinuous character of these transitions relies on the analysis of the full probability distribution $P(\mu)$. For sufficiently large lattices, $P(\mu)$ develops a clear bimodal structure with disjoint maxima, which is indicative of the coexistence of structurally distinct absorbing states across independent simulation runs. Our refined analysis reveals that for $F=2$, the system undergoes a genuinely hybrid phase transition: the order parameter exhibits a remarkably small but finite latent jump ($\mu_c \approx 0.089$) at the critical threshold, yet it scales toward the transition point from the fragmented regime via a non-analytical power-law with a mean-field exponent $\beta \approx 1/2$. It is worth noting that while the two-dimensional Potts model with five states faces severe numerical masking due to its weak first-order nature, the hybrid character of the transition for $F=2$ in our model makes the numerical distinction of the discontinuity an even more formidable and elusive task. Conversely, for $F=3$, the increased dimensionality of the trait space suppresses local fluctuations, turning the system into a traditional, non-hybrid first-order transition where the order parameter behaves in a strictly analytical manner before its abrupt collapse. We apply this methodology to the discrete Poisson variant (detailed in \ref{appA}) and confirm the previous conclusions of a continuous transition for $F=2$ and a discontinuous, non-hybrid transition for $F=3$.

\section{The Continuous-Trait Axelrod Model}\label{sec:mod}

We consider the continuous-trait version of the Axelrod model recently introduced in Ref.~\cite{Campos_2025}. In this formulation, each agent $i = 1, \ldots, L^2$ is located on a square lattice of linear size $L$ with periodic boundary conditions. The cultural state of an agent $i$  is defined by a vector of $F$ traits, $\mathbf{x}_i = (x_i^1, \ldots, x_i^F)$, where each component $x_i^k$ is a continuous variable sampled from a uniform distribution in the interval $[0,1]$.

The interaction between neighboring agents $i$ and $j$ is governed by the principle of homophily, quantified by the similarity $p_{ij}$. Two traits are considered similar if their distance is within a threshold $d \in [0,1]$. Formally, $p_{ij}$ is the fraction of such traits,
\begin{equation}\label{pij}
p_{ij}  = \frac{1}{F} \sum_{k=1}^F \Theta \left (d - |x_i^k - x_j^k|  \right ),
\end{equation}
where $\Theta(z)$ is the Heaviside step function. The dynamics follow a standard stochastic protocol: at each time step, a focal agent $i$ and one of its four nearest neighbors $j$ are selected at random. An interaction occurs with probability $p_{ij}$. If successful, agent $i$ randomly selects a trait $k$ from the subset of non-similar traits ($|x_i^k - x_j^k| > d$) and copies it from neighbor $j$, such that $x_i^k = x_j^k$. Unlike the noisy process explored in Ref.~\cite{Campos_2025} (see also \cite{Mora_2025}), here we focus exclusively on the limit of perfect copying, which allows for a sharper characterization of the absorbing states.

The system eventually reaches an absorbing configuration when $p_{ij} \in \{0, 1\}$ for all neighboring pairs. These states are characterized by the emergence of cultural domains, defined as connected components of a graph where edges exist only between neighbors $i$ and $j$ that satisfy $|x_i^k - x_j^k| \leq d$ for all $k=1, \ldots, F$ \cite{Newman_2018}. Importantly, this definition implies that while adjacent agents within a domain are similar, non-neighboring agents in the same domain  may have traits that differ by more than $d$.

Upon reaching an absorbing state, we quantify cultural fragmentation through two order parameters: the density of cultural domains ($\mu$), defined as the ratio of the number of domains to the total number of agents, and the fraction of agents in the largest domain ($\rho$). For comparison, in standard site percolation on a 2D lattice, $\mu$ is a continuous and differentiable function that remains non-zero at the critical point, whereas $\rho$ undergoes a continuous transition governed by the exponent $\beta = 5/36$ \cite{Stauffer_1992}.

Unlike the Poisson variant of the Axelrod model~\cite{Castellano_2000, Vilone_2002}, where both order parameters are believed to exhibit continuous transitions for $F=2$~\cite{Peres_2015, Reia_2016}, the continuous-trait variant yields an unmistakably discontinuous transition for the fraction of sites in the largest domain, $\rho$, as evidenced by the characteristic crossing of the $\rho$ versus $d$ curves for different lattice sizes $L$~\cite{Campos_2025}. While previous analyses relying on ensemble averages suggested that the domain density $\mu$ undergoes a continuous transition for both $F=2$ and $F=3$, such macroscopic averages can be deceptive in the presence of strong fluctuations or bimodality. Here, we resolve this ambiguity by shifting our focus from simple mean values to the median and the full probability distributions of the order parameters, thereby providing a more comprehensive characterization of the model's critical behavior. As will be shown below, this approach reveals that the transition for the domain density $\mu$ is, in fact, discontinuous in both cases.

\section{The hybrid transition for $F=2$}\label{sec:CF2}

Let us focus first on the more challenging case of $F=2$. As previously noted, Ref.~\cite{Campos_2025} characterized the transition for the domain density as continuous, based on the smooth vanishing of its mean value $\bar{\mu}$ with an estimated exponent $\beta \approx 1/3$. In that work, the critical threshold $d_c \approx 0.0784$ was identified through the intersection of the coefficient of variation (CV) of $\mu$ for different lattice sizes. By focusing instead on the median $\tilde{\mu}$ and significantly extending the computational scope to larger lattice sizes up to $L=1600$, we are able to re-examine this scenario and resolve the true nature of the transition.

\begin{figure}[ht] 
\centering
 \includegraphics[width=1\columnwidth]{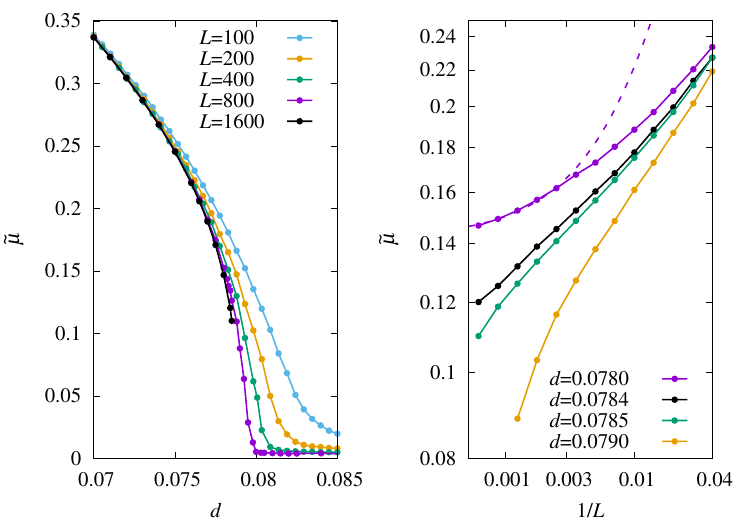}  
\caption{Median $\tilde{\mu}$ of the domain density for $F=2$ as a function of the threshold $d$ for lattice sizes $L=100, 200, 400, 800$, and $1600$ (left panel), and as a function of the reciprocal of the lattice size $1/L$ for thresholds $d = 0.0780,  0.0784,  0.0785$, and $0.0790$ (right panel). For $L=1600$, the data were collected up to $d=0.0785$. The dashed curve represents the fit $\tilde{\mu} = \tilde{\mu}_\infty + b/L$ to the data for $d=0.0780$. The median exhibits a continuous behavior across all simulated system sizes.}   
\label{fig:F2tt}  
\end{figure}

Figure~\ref{fig:F2tt} illustrates the functional dependence of the median $\tilde{\mu}$ on both the threshold $d$ and the reciprocal of the lattice size $1/L$ within the critical region. For $L=1600$, the data are presented only up to $d=0.0785$; extending the simulations further into the homogeneous regime is not only computationally prohibitive due to extremely slow convergence times but also irrelevant, as our primary interest lies in fitting the order parameter $\tilde{\mu}$ within the region where it remains nonzero. The median varies smoothly and monotonically with $d$ and $L$, exhibiting neither curve crossings nor abrupt jumps for different system sizes. For $d < d_c$, we expect $\tilde{\mu}$ to approach a nonzero value in the thermodynamic limit ($L \to \infty$); consequently, the curve of $\tilde{\mu}$ versus $1/L$ on a log-log scale must be convex. To obtain the asymptotic thermodynamic median value, we fit the data using the ansatz $\tilde{\mu} = \tilde{\mu}_\infty + b/L$, where $\tilde{\mu}_\infty$ and $b$ are fit parameters that depend on $d < d_c$. In particular, we find $\tilde{\mu}_\infty \approx 0.142$ for $d=0.0780$. 

Conversely, for $d > d_c$, $\tilde{\mu} \to 0$ as $L \to \infty$, meaning that the log-log curve of $\tilde{\mu}$ versus $1/L$ must be concave. These geometric criteria hold true regardless of whether the transition is continuous or discontinuous, providing a highly reliable method for estimating the critical point. Indeed, the scaling behavior shown in the right panel of the figure yields the estimate $d_c \in (0.0784, 0.0785)$. Narrowing this interval further would require simulating even larger lattice sizes, which is currently computationally unfeasible. Nevertheless, the very slight convexity of the curve for $d=0.0784$ signals its extreme proximity to the critical threshold; hence, we shall adopt the estimate $d_c \approx 0.0784$ given in Ref.~\cite{Campos_2025}.

A word is in order regarding the number of independent runs (samples) used for the estimates of the median $\tilde{\mu}$ presented in Fig.~\ref{fig:F2tt}. The minimum number of samples used was $5000$ for the largest lattice sizes ($L=1131$ and $L=1600$) above $d_c$; for all other sizes, we typically employed $10^4$ samples, which guarantees precise estimates of the median. Below $d_c$, the absorption dynamics converge very rapidly, allowing us to easily collect $10^4$ samples for all lattice sizes. Furthermore, we remark that increasing $L$ systematically decreases the estimate of the median; consequently, the median for any finite lattice size acts as an upper bound to the median of the infinitely large system.

\begin{figure}[ht] 
\centering
 \includegraphics[width=1\columnwidth]{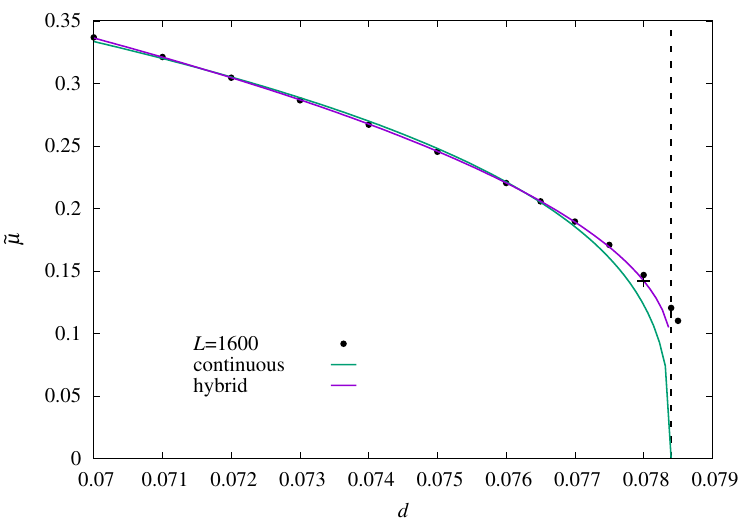}  
\caption{Median $\tilde{\mu}$ of the domain density for $F=2$ as a function of the threshold $d \leq 0.0785$ for the largest simulated lattice size ($L=1600$). The solid curves display the best fits obtained under the continuous and hybrid transition hypotheses. The vertical dashed line indicates the independent estimate of the critical threshold, $d_c = 0.0784$, whereas the symbol $+$ indicates the asymptotic thermodynamic median value $\tilde{\mu}_\infty \approx 0.142$ previously extrapolated for $d=0.0780$.}   
\label{fig:F2cd}  
\end{figure}

Let us assume, as done in Ref.~\cite{Campos_2025} for the mean domain density, that the median $\tilde{\mu}$ vanishes continuously at $d_c$. Under this assumption, we can formulate the scaling ansatz $\tilde{\mu} = \mathcal{A} (d_c-d)^\beta$, which is expected to hold in the thermodynamic limit within the immediate vicinity below the critical threshold.  Since our data for different system sizes overlap perfectly at $d \leq 0.0770$, indicating the absence of finite-size corrections, the results for $L=1600$ provide an accurate approximation of the thermodynamic limit. We thus fit the ansatz to these points by adjusting the amplitude $\mathcal{A}$ and the critical exponent $\beta$.The resulting scaling curve is shown in Fig.~\ref{fig:F2cd}. Although the coefficient of determination for this fit is remarkably high ($R^2 = 0.9991$), yielding best-fit parameters $\mathcal{A}=1.60 \pm 0.05$ and $\beta = 0.328 \pm 0.006$, a closer inspection of the figure raises two major red flags. 

First, the fitted curve systematically overshoots the data points that represent the true thermodynamic limit. This forced upward bending of the curve stems from the fixed estimate of $d_c = 0.0784$; indeed, this artifice would be even more accentuated if we were to use a lower estimate of $d_c$. We note that an ostensibly flawless fit can be achieved if $d_c$ is also allowed to vary freely, which yields a higher critical threshold of $d_c = 0.0793$ (with the amplitude and exponent changing accordingly). However, treating $d_c$ as a free parameter directly contradicts our independent, robust geometric estimate of the critical threshold obtained from the scaling analysis in the right panel of Fig.~\ref{fig:F2tt}. Second, the continuous-transition fit fails spectacularly to capture the asymptotic thermodynamic median value $\tilde{\mu}_\infty \approx 0.142$ previously extrapolated for $d=0.0780$.

Since the assumption of a continuous transition raises structural suspicions, let us consider the alternative scenario in which  $\tilde{\mu}$ is discontinuous at $d_c = 0.0784$. Under this framework, we propose the scaling ansatz 
\begin{equation}\label{hybrid}
\tilde{\mu} = 
\begin{cases} 
\mu_c + \mathcal{A} (d_c-d)^\beta & \text{for } d \leq d_c, \\ 
0 & \text{for } d > d_c,  
\end{cases}
\end{equation}
where the additional fit parameter $\mu_c$ represents the finite jump (latent density) of the order parameter at the critical threshold. The resulting, virtually flawless fit is also displayed in Fig.~\ref{fig:F2cd}, yielding the best-fit parameters $\mu_c = 0.089 \pm 0.003$, $\mathcal{A} = 2.7 \pm 0.1$, and $\beta = 0.50 \pm 0.01$. The convergence toward $\beta \approx 0.5$ alongside a non-zero latent jump strongly suggests that the continuous-trait variant for $F=2$ undergoes a hybrid phase transition. This fascinating class of transitions combines characteristics of both first- and second-order behaviors simultaneously. Remarkably, our estimated exponent is in perfect agreement with the  value $\beta = 1/2$ found in classic examples of hybrid transitions, such as $k$-core percolation~\cite{Dorogovtsev_2006, Baxter_2015} and the percolation of interdependent networks~\cite{Lee_2016}.

\begin{figure}[ht] 
\centering
 \includegraphics[width=1\columnwidth]{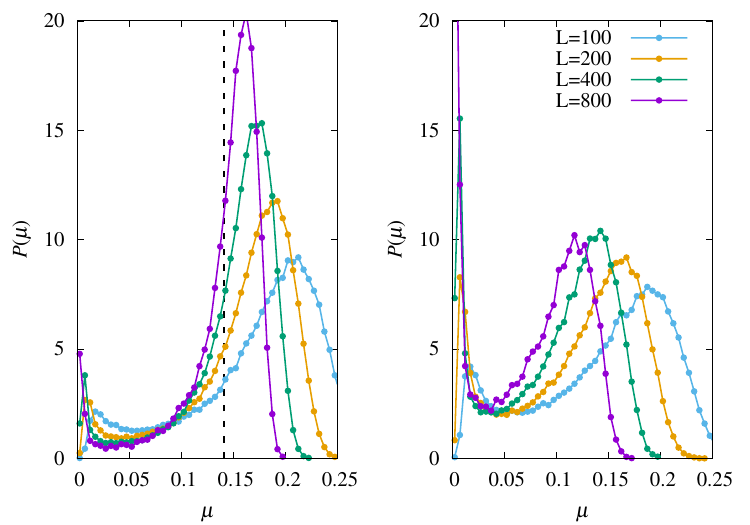}  
\caption{Distribution of the domain density $P(\mu)$ for $F=2$ at $d=0.0780$ (left) and $d=0.0790$ (right) for lattice sizes $L=100, 200, 400$, and $800$. The bimodal structure near the transition point reflects the competition between fragmented and homogeneous absorbing states. In the left panel, the vertical dashed line at $\mu \approx 0.142$ indicates the location of the thermodynamic limit extrapolation for the median. In both panels, the vertical scale is truncated at $P(\mu) = 20$ to better visualize the behavior of the distribution at intermediate and high values of $\mu$. Data for $L=1600$ are omitted because the sample size is insufficient to fully resolve the distribution shape without noise.}   
\label{fig:F2dmu_1}  
\end{figure}

Although the superior quality of the fit achieved under the hybrid-transition hypothesis~\eqref{hybrid} provides compelling macroscopic evidence, a deeper understanding of the transition requires examining the underlying microscopic statistics. Analyzing the full domain density distribution $P(\mu)$ allows us to independently substantiate this scenario and uncover the microstructural signatures of the critical region. To this end, Fig.~\ref{fig:F2dmu_1} displays $P(\mu)$ within both the fragmented regime at $d=0.0780$ and the homogeneous regime at $d=0.0790$, where the competing nature of the absorbing configurations becomes directly visible.

Indeed, a bimodal-like structure is evident in the distributions for both regimes, yet it differs significantly from the sharp split observed in true discontinuous transitions, such as that exhibited for the $F=3$ case, discussed in Sect.~\ref{sec:CF3}. Notably, the peak corresponding to the fragmented configurations is not stationary at a fixed nonzero value of $\mu$ but shifts continuously toward $\mu=0$ as $L$ increases. This finite-size behavior poses a formidable numerical challenge: distinguishing whether the two peaks eventually merge into a single, connected probability mass in the thermodynamic limit, or whether the statistical ensemble dynamically splits into two mutually exclusive absorbing regimes.

The left panel of Fig.~\ref{fig:F2dmu_1} ($d=0.0780$) shows that the valley between the two peaks deepens as $L$ increases, while the peak associated with the fragmented regime grows in height and shifts toward the left. This deepening of the minimum between the peaks is a characteristic finite-size signature of an impending discontinuity as the thermodynamic limit is approached. The intersection point between the density distributions for different $L$ supports the  existence of the second peak in the thermodynamic limit, located around the extrapolated value of the median $\tilde{\mu}_\infty \approx 0.142$. 

Conversely, the right panel of Fig.~\ref{fig:F2dmu_1} ($d=0.0790$) illustrates a case where the valley does not deepen with increasing $L$. Instead, the peak corresponding to the fragmented configurations shifts toward $\mu=0$ as $L$ increases, indicating that the two regimes eventually merge in the thermodynamic limit. This behavior implies the absolute dominance of the homogeneous regime ($\mu=0$) at this value of $d$. This asymmetric behavior provides a clear microscopic confirmation of the hybrid phase transition hypothesis~\eqref{hybrid}; the non-analytical approach of the order parameter toward $d_c$ abruptly disrupts the mechanism safeguarding the finite-size stability of the fragmented peak, triggering the   jump of the order parameter to zero.

It is worth noting that this hybrid behavior stems directly from the continuous nature of the cultural traits. As detailed in \ref{appA}, when the same analysis is applied to the Poisson variant of the Axelrod model (where traits remain discrete), the transition for $F=2$ is continuous, governed by a standard power-law scaling that lacks any latent jump.
\section{The discontinuous transition for $F=3$}\label{sec:CF3}

\begin{figure}[ht] 
\centering
 \includegraphics[width=1\columnwidth]{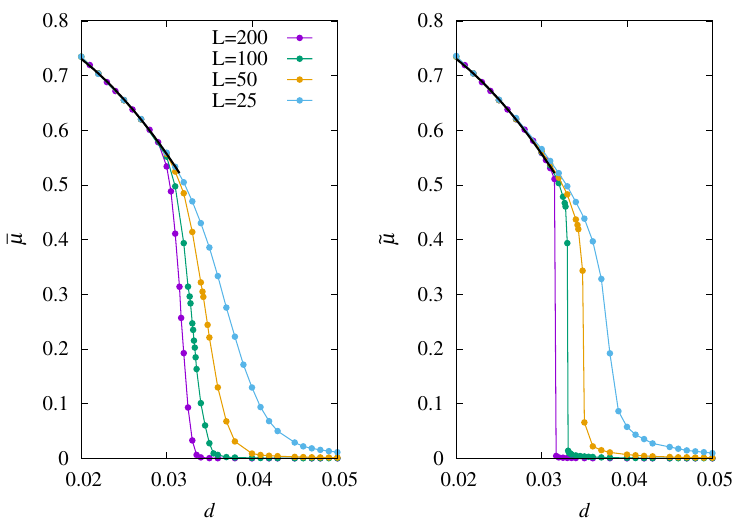}  
\caption{Mean $\bar{\mu}$ (left panel) and median $\tilde{\mu}$ (right panel) of the density of domains as a function of the threshold $d$ for $F=3$ and lattice sizes $L=25, 50, 100$, and $200$. While the mean appears smooth and continuous for these finite sizes, the median exhibits a jump for $L \ge 50$ at an effective threshold that shifts toward the thermodynamic value $d_c \approx 0.031$ as $L$ increases. The solid curves represent the analytical polynomial fit for the peak position $\mu_0(d)$, defined in Eq.~\eqref{eq:toy}, taking the regular form $\mu_0(d) = \mu_c + \mathcal{A}(d_c-d) + \mathcal{B} (d_c-d)^2$ for the $L=200$ data, which yields $\mu_c = 0.533 \pm 0.001$, $\mathcal{A}= 23.5 \pm 0.5$, and $\mathcal{B}=-515 \pm 55$. Remarkably, this single analytical function successfully captures the regular background variation of both the mean (left)  and the median (right)   before the abrupt collapse.}   
\label{fig:F3bt}  
\end{figure}

Here, we analyze the case $F=3$, where the discontinuous nature of the transition for the domain density $\mu$ is more pronounced, though still challenging to characterize. In Fig.~\ref{fig:F3bt}, we present both the mean ($\bar{\mu}$) and the median ($\tilde{\mu}$) of the density of domains, obtained from a large number of independent simulations (ranging from $10^4$ to $10^5$ realizations). The critical threshold is estimated at $d_c \approx 0.031$, based on the crossings of the $\rho$ versus $d$ curves for different lattice sizes $L$ (not shown). Nevertheless, the abrupt jump exhibited by the median for $L=200$ at $d = 0.0315$ already provides a remarkably close estimate of this critical point.

The absence of crossings in the $\bar{\mu}$ curves across different lattice sizes $L$ (see left panel of Fig.~\ref{fig:F3bt}) makes it difficult to determine the order of the transition based solely on the mean. Indeed, a very small critical exponent $\beta$---such as that of 2D site percolation~\cite{Stauffer_1992}---could produce a smooth yet sharp variation within the critical region, which inadvertently led Ref.~\cite{Campos_2025} to conclude that the domain density transition was continuous. Conversely, the behavior of the median $\tilde{\mu}$ is completely unambiguous: it exhibits a distinct jump for $L \ge 50$, with the magnitude of the discontinuity systematically increasing with $L$ (see right panel of Fig.~\ref{fig:F3bt}). This clear macroscopic jump directly signals a discontinuous transition, even though $\tilde{\mu}$ still appears continuous for the smallest size analyzed ($L=25$).

%-----------------------------------------------------
%-----------------------------------------------------

\begin{figure}[ht] 
\centering
 \includegraphics[width=1\columnwidth]{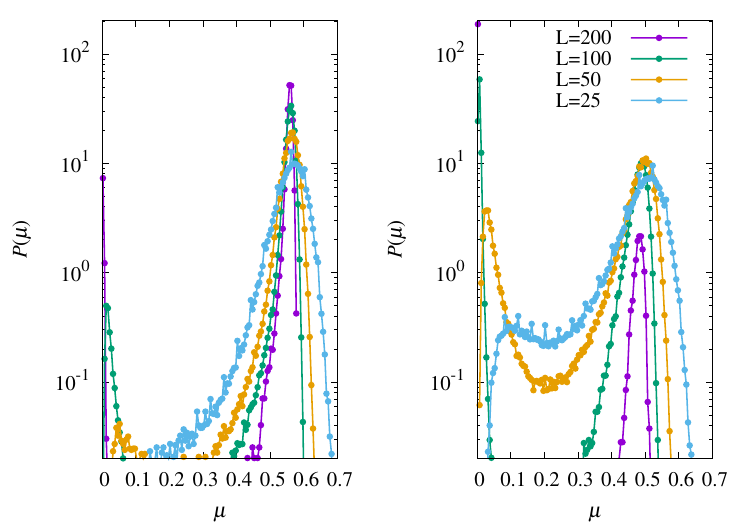}  
\caption{Distribution of the density of domains $P(\mu)$ for $F=3$ at $d=0.030$ (left panel) and $d=0.033$ (right panel) for lattices of linear sizes $L=25, 50, 100$, and $200$. The emergence of a bimodal distribution for large $L$ near the transition point illustrates the competition between fragmented and homogeneous absorbing states.}   
\label{fig:F3dmu}  
\end{figure}

The origin of this discrepancy between the mean and the median can be understood by examining the full distribution of the density of domains, $P(\mu)$, rather than relying on summary statistics. Figure~\ref{fig:F3dmu} shows $P(\mu)$ in the vicinity of the transition ($d=0.030$ and $d=0.033$). For large $L$, the distribution is clearly bimodal, with two sharp, separate peaks representing two distinct classes of absorbing states: some realizations of the dynamics reach a fragmented configuration, while others converge to a culturally homogeneous state. 

We note that for $d=0.030$, the height of both peaks increases with $L$. This occurs because, as the system size increases, finite-size fluctuations decrease, and the probability mass that was previously spread across intermediate values of $\mu$ concentrates into these two dominant absorbing regimes. In the case of $d=0.030$, the peak associated with the fragmented configurations eventually gains more weight as $L$ grows, as it represents the stable absorbing regime for this threshold value. This competition between two increasingly sharp, isolated peaks is the hallmark of a first-order transition, where the change in the order parameter is driven by a split of the statistical ensemble and the transfer of probability weight from one discrete macrostate to another.

%-----------------------------------------------------
%-----------------------------------------------------

To clarify how this bimodality affects the order parameters and to elucidate the microscopic mechanism that distinguishes a traditional first-order transition from a hybrid one, we model this scenario through a simple toy model. We represent the distribution $P(\mu)$ as the sum of two delta functions:
\begin{equation}\label{eq:toy}
P(\mu) = a(L,d) \delta(\mu) + b(L,d) \delta \left[ \mu - \mu_0(d) \right],
\end{equation}
where $\mu_0(d)$ represents the position of the fragmented peak, which depends smoothly on the threshold $d$. The weights $a(L,d)$ and $b(L,d)$ denote the fractions of independent realizations ending in the homogeneous and fragmented states, respectively ($a+b=1$). Inspired by finite-size scaling theory, these weights can be modeled by the sigmoidal form:
\begin{equation}
b(L,d) = \frac{1}{1 + \exp\left[ (d - d_c) L^{1/\nu} \right]},
\end{equation}
where $d_c$ is the critical threshold and $\nu$ is an effective correlation length exponent. 

For any finite lattice size $L$, the weight $b(L,d)$ is a smooth, continuous function of $d$. Consequently, the mean density of domains, given by $\bar{\mu}(d) = \mu_0(d) b(L,d)$, appears continuous in numerical simulations for finite systems, masking the underlying discontinuity. Conversely, the median $\tilde{\mu}$ tracks the position of the dominant peak; it abruptly jumps from $\mu_0(d)$ to $0$ at the exact point where $b(L,d)$ crosses $0.5$ (i.e., at $d=d_c$), explaining why the discontinuity is immediately visible in the median data for $L \ge 50$. In the thermodynamic limit ($L \to \infty$), $b(L,d)$ converges to the Heaviside step function $\Theta(d_c - d)$, and the sharp discontinuity is recovered for both estimators, yielding $\bar{\mu}(d) = \mu_0(d)\Theta(d_c - d)$.

This toy model allows us to distinguish between a hybrid transition and a traditional first-order one by examining the mathematical nature of $\mu_0(d)$ at the critical point. While a hybrid transition (such as the one observed for $F=2$) requires $\mu_0(d)$ to approach the critical threshold non-analytically via a power-law with an exponent $\beta < 1$, our data for $F=3$ reveals a strictly analytical behavior. As shown by the solid curves in Fig.~\ref{fig:F3bt}, the fragmented regime is perfectly described by a regular polynomial expansion, $\mu_0(d) = \mu_c + \mathcal{A}(d_c-d) + \mathcal{B} (d_c-d)^2$, with no divergence in its derivatives at $d_c$. Remarkably, because the median directly tracks $\mu_0(d)$ for $d \le d_c$ and the mean converges to $\mu_0(d)$ as $b(L,d) \to 1$, this single analytical fit successfully characterizes the smooth background variation of both statistical estimators before the collapse. This proves that the transition for $F=3$ lacks critical fluctuations in the order parameter, behaving as a classical, non-hybrid discontinuous transition driven entirely by the transfer of probability weight between two stable, localized peaks.

This shift toward a traditional first-order transition driven by the increased dimensionality $F$ of the trait space is not exclusive to the continuous-trait model. In  \ref{appA}, we show that the Poisson variant for $F=3$ undergoes a fundamentally identical discontinuous transition, highlighting that a larger trait space suppresses finite-size fluctuations and promotes phase separation, regardless of whether the underlying traits are continuous or discrete.

\section{Discussion}\label{sec:dis}

In this work, we have provided a comprehensive re-examination of the phase transitions in the continuous-trait Axelrod model, uncovering a far more intricate scenario than previously reported. In particular, our results allow us to refine the preliminary characterization offered in Ref.~\cite{Campos_2025}. In that study, the term "hybrid transition" was employed in an unconventional sense: it described a mixed scenario where the domain density $\mu$ vanished continuously, whereas the relative size of the giant component $\rho$ exhibited a discontinuity. By shifting our methodological approach toward the median $\tilde{\mu}$ and analyzing the full probability distributions $P(\mu)$ across significantly larger lattice sizes, we have resolved the subtle finite-size effects that masked the true nature of these transitions, offering a corrected  perspective in line with modern statistical physics definitions. 

For the $F=2$ case, the apparent continuity of the domain density reported in Ref.~\cite{Campos_2025} stemmed from a small latent jump ($\mu_c \approx 0.089$), which is notoriously difficult to detect using standard ensemble averages due to strong finite-size fluctuations within the critical region. Our scaling analysis of the median effectively unveiled this hidden discontinuity. More importantly, it revealed that the transition is truly hybrid in the contemporary sense of the term: the order parameter jumps to zero  at the critical threshold $d_c$, yet it approaches this point from below via a non-analytical power-law characterized by an exponent $\beta \approx 1/2$. This places the continuous-trait Axelrod model for $F=2$ in the same fascinating class of non-equilibrium phenomena as $k$-core percolation~\cite{Dorogovtsev_2006, Baxter_2015} and interdependent networks~\cite{Lee_2016}, where second-order fluctuations are intricately intertwined with a first-order collapse.

Similarly, our current framework successfully clarifies the behavior of the system for $F=3$. While Ref.~\cite{Campos_2025} inadvertently concluded that the domain density transition remained continuous for higher trait dimensions, the present analysis shows that the $F=3$ case undergoes a traditional, non-hybrid discontinuous transition. Here,  the order parameter approaches the critical threshold in a strictly analytical manner—perfectly captured by a regular polynomial Taylor expansion—and the abrupt collapse of the order parameter is driven entirely by the transfer of probability weight between two isolated, mutually exclusive absorbing regimes. 

To put these findings into perspective, we compared our continuous-trait results with the discrete-trait Poisson variant of the Axelrod model~\cite{Castellano_2000, Vilone_2002} (detailed in \ref{appA}). For $F=2$, this comparison reveals a notable contrast: while the continuous-trait model introduces a hybrid transition, the Poisson variant undergoes a  continuous transition. Our finite-size scaling analysis yields a critical exponent $\beta \approx 0.60$ and a threshold $q_c = 3.10$, in agreement with the ensemble-average results reported in Ref.~\cite{Peres_2015}. Microstochastically, although the domain density distribution $P(\mu)$ displays a bimodal-like structure due to finite-size fluctuations, the peaks remain connected by a robust probability mass that does not vanish as $L$ increases, confirming the absence of mutually exclusive absorbing regimes. In contrast, for $F=3$, the Poisson variant mirrors the continuous-trait model by undergoing a purely discontinuous transition. This convergence of results shows that while the dimensionality of the trait space ($F$) universally dictates whether the transition is continuous or discontinuous, the mathematical nature of the traits themselves (continuous versus discrete) plays a pivotal role in shaping the hybrid characteristics of the collapse.

From a physical perspective, the  difference between the continuous phase transition in the Poisson variant and the hybrid transition in the continuous-trait model for $F=2$ can be understood through the lenses of local fluctuations and effective domain connectivity. In the continuous-trait formulation, the interaction rule is governed by a metric threshold $d$ between immediate neighbors.  This allows members belonging to the same macroscopic cultural domain to exhibit traits whose mutual distance is significantly larger than $d$, provided they are linked through a continuous path of locally compatible neighbors. This cumulative stretching of the trait space enables the emergence of larger, more coordinated cultural domains than those formed via exact discrete matches. Mechanistically, these larger configurations act to suppress local statistical fluctuations across the lattice, effectively pushing the system toward a mean-field-like behavior where discontinuous jumps are favored~\cite{Stanley_1971}.

In contrast, the discrete nature of the Poisson variant preserves strong local fluctuations, as cultural alignment relies on exact integer overlaps, restricting domain growth. In physics, such local fluctuations are well known to smooth out discontinuities, explaining why the Poisson variant for $F=2$ softens into a strictly continuous transition. For the continuous-trait model, however, the competition between the fluctuation-suppressing nature of these expanded geometric domains and the structural constraints of the 2D lattice results in a compromise: a hybrid transition that retains a finite latent jump ($\mu_c$) while scaling toward the threshold via a non-analytical mean-field exponent $\beta \approx 1/2$.

A natural next step is to investigate the robustness of the hybrid transition for $F=2$ under alternative structural and dynamical modifications. On one hand, moving toward complex or multilayer networks, where long-range connections suppress local fluctuations, is expected to destabilize the critical scaling and potentially drive the system toward fully discontinuous, non-hybrid collapses~\cite{Reia_2016,Barrat_2000}. On the other hand, the hybrid nature could be softened into a strictly continuous transition by introducing mechanisms that enhance local fluctuations or relax the metric rigidity of the interaction rule. This could be achieved either by implementing a smooth, probabilistic interaction profile instead of a sharp bounded-confidence threshold~\cite{Sanctis_2009}, or by incorporating agent heterogeneity through a distribution of individual tolerance limits~\cite{Lorenz_2010}---a mechanism closely tied to threshold models of collective behavior~\cite{Granovetter_1978, Fontanari_2026}. Unveiling how these competing structural and dynamical ingredients expand or suppress the hybrid regime remains a fertile ground to characterize how consensus and fragmentation compete in continuously evolving societies.

\section*{Acknowledgments}

JFF is partially supported by  Conselho Nacional de Desenvolvimento Cient\'{\i}fico e Tecnol\'ogico  grant number 300200/2026-9.  
PRAC was partially supported by Conselho Nacional de Desenvolvimento Cient\'{\i}fico e Tecnol\'ogico (CNPq) under Grant No. 301795/2022-3,   Funda\c{c}\~ao de Amparo \`a Ci\^encia e Tecnologia do Estado de Pernambuco (FACEPE), Grant Number APQ-1129-1.05/24, The National Institute of Science and Technology (INCT) in Ecology, Evolution, and Biodiversity Conservation funded by CNPq (grant 409197/2024-6 and 384712/2025-8) and FAPEG (grant 201810267000023), and National Institute of Science and Technology in Innovative Research in Health Sciences: from Nanotechnology to Artificial Intelligence sponsored by CNPq (grant no. 408417/2024-2) and FAPESP (grant no. 2025/26818-7).

%section*{Declarations}

%\subsection*{Funding} Conselho Nacional de Desenvolvimento Cient\'{\i}fico e Tecnol\'ogico and Funda\c{c}\~ao de Amparo \`a Ci\^encia e Tecnologia do Estado de Pernambuco.
%subsection*{Conflict of interest} The authors declare that they have no Conflict of interest.
%\item Ethics approval and consent to participate Not applicable.
%\item Consent for publication Not applicable.
%\item Data availability Not applicable.
%\item Materials availability Not applicable.
%\item Code availability Not applicable.
%\subsection*{Author contribution}  All authors contributed equally to this work.

\newpage

\appendix

\section{The Poisson variant of the Axelrod model}\label{appA}
\renewcommand{\theequation}{A.\arabic{equation}}
\setcounter{equation}{0}
\setcounter{figure}{0}

%-----------------------------------------------------
\begin{figure}[ht] 
\centering
 \includegraphics[width=1\columnwidth]{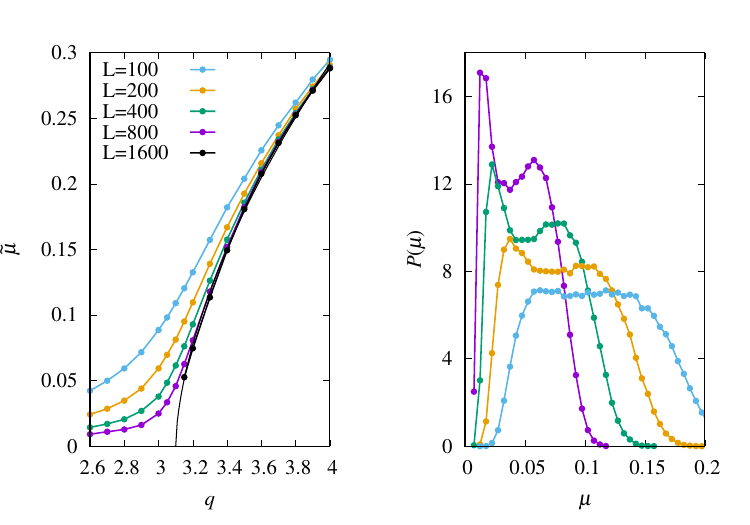}  
\caption{(Left) Median domain density $\tilde{\mu}$ for $F=2$ vs. $q$; the solid curve is a power-law fit $\tilde{\mu} \propto (q-q_c)^\beta$ for $L=1600$, yielding $\beta = 0.60 \pm 0.01$ with $q_c=3.10$ fixed. (Right) Probability distribution $P(\mu)$ for $F=2$ at $q=3.10$, showing a connected probability mass despite finite-size fluctuations. Data for $L=1600$ are omitted in the right panel because the sample size is insufficient to fully resolve the distribution shape without noise, although it remains adequate for estimating the median.}   
\label{fig:F2Pdmu}  
\end{figure}
%-----------------------------------------------------

To assess the robustness of our methodological framework, we analyze the Poisson variant of the Axelrod model~\cite{Castellano_2000, Vilone_2002}. In this formulation, each trait $\sigma_i^k$ of an agent's cultural vector $\mathbf{\sigma}_i = (\sigma_i^1, \dots, \sigma_i^F)$ is a non-negative integer sampled from a Poisson distribution with parameter $q \in (0, \infty)$, where the cultural similarity $p_{ij}$ between neighbors is defined by the overlap $p_{ij} = \frac{1}{F} \sum_{k=1}^F \delta(\sigma_i^k, \sigma_j^k)$. Here, $q$ regulates the initial cultural diversity, playing a role inverse to the threshold $d$ in the continuous-trait model. For $F=2$, this variant is known to undergo a continuous phase transition at $q_c \approx 3.10$~\cite{Castellano_2000, Peres_2015, Reia_2016}, whereas for $F=3$, early studies suggested a discontinuous transition based on qualitative domain size distributions~\cite{Castellano_2000}.

For the $F=2$ case, our simulations with large lattice sizes up to $L=1600$ firmly consolidate the continuous transition scenario (Fig.~\ref{fig:F2Pdmu}). A two-parameter power-law fit of the median $\tilde{\mu} = \mathcal{A}(q-q_c)^\beta$ yields a critical exponent $\beta = 0.60 \pm 0.01$ (with $q_c = 3.10$ fixed via concavity analysis), in agreement with the $\beta \approx 0.67$ reported for ensemble averages in Ref.~\cite{Peres_2015}. This continuous nature is microstochastically corroborated by the full probability distribution $P(\mu)$ near the threshold. Although a bimodal-like structure emerges due to strong finite-size fluctuations, the underlying probability mass remains robustly connected, and the valley separating the peaks does not drop to zero as $L$ increases, effectively ruling out a sharp  split into two disjoint absorbing phases in the large $L$ limit.

%-----------------------------------------------------
\begin{figure}[t] 
\centering
 \includegraphics[width=1\columnwidth]{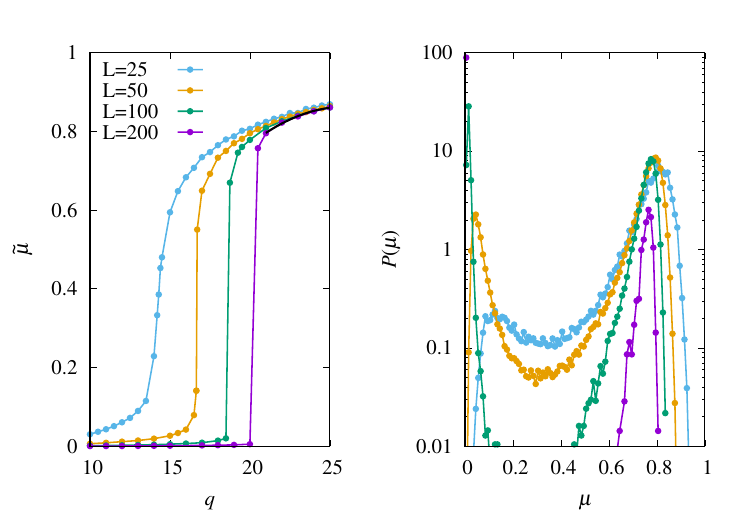}  
\caption{(Left) Median domain density $\tilde{\mu}$  for $F=3$ as a function of the Poisson parameter $q$. While the transition appears smooth for $L=25$, a discontinuity emerges as the system size increases, with the jump occurring at $q=q_c \approx 24$ in the thermodynamic limit.  The solid curve represents the analytical polynomial fit $\tilde{\mu} = \tilde{\mu}_c + \mathcal{A}(q-q_c) + \mathcal{B} (q-q_c)^2$ for the $L=200$ data, which yields $\tilde{\mu}_c = 0.851 \pm 0.001$, $\mathcal{A}= 0.011\pm 0.001$, and $\mathcal{B}=-0.0026 \pm 0.0005$.  (Right) Probability distribution $P(\mu)$ for $q=19$. The bimodality illustrates the competition between a metastable fragmented state and the stable homogeneous state ($\mu \approx 0$). As $L$ increases, the peak corresponding to the homogeneous regime becomes dominant. }   
\label{fig:F3Pdmu}  
\end{figure}
%-----------------------------------------------------

The results for $F=3$ reveal a completely different picture that closely mirrors our findings for the continuous-trait model. As shown in Fig.~\ref{fig:F3Pdmu}, while the ensemble mean varies smoothly (not shown), the median $\tilde{\mu}$ exhibits a macroscopic jump for $L \ge 50$ at an effective threshold that systematically shifts toward the thermodynamic critical value $q_c \approx 24$ as $L$ increases. This discontinuity is driven by the emergence of two disjoint and well-separated peaks in the distribution $P(\mu)$, indicating a clear split of the statistical ensemble into two mutually exclusive absorbing regimes. As $L$ increases, the probability weight is transferred between these localized states, and the homogeneous peak at $\mu \approx 0$ eventually dominates.

\end{document}